\documentclass[showpacs,amsmath,amssymb]{revtex4}
\usepackage{amssymb}
\usepackage{graphicx}

\begin{document}

\title{Magnetoresistance in organic semiconductors: including pair correlations in the kinetic equations for hopping transport}
\author{A. V. Shumilin$^1$, V.V. Kabanov$^2$, V.I. Dediu$^3$}

\affiliation{$^1$ Ioffe Institute, 194021 St.-Petersburg, Russia}
\affiliation{$^2$ Department for Complex Matter, Jozef Stefan Institute, 1001 Ljubljana, Slovenia}
\affiliation{$^3$ CNR - ISMN, via Gobetti 101, 40129 Bologna, Italy}

\begin{abstract}
 We derive the kinetic equations for polaron hopping in organics that explicitly take into account the
 double occupation possibility and pair intersite correlations. The equations include simplified phenomenological spin dynamics and provide a self-consistent framework for the description of the bipolaron mechanism of the organic magnetoresistance. At low applied voltages the equations can be reduced to effective resistor network that generalizes the Miller-Abrahams network and includes the effect of spin relaxation on the system resistivity. Our theory discloses the close relationship between the organic magnetoresistance and the intersite correlations. Moreover, in the absence of correlations, as in ordered system with zero Hubbard energy, the magnetoresistance vanishes.
\end{abstract}


\maketitle
\section{Introduction}\label{sec1}

The transport properties of organic materials represent an interesting and fast-developing research field. While a number of basic properties, especially those related to light emitting devices\cite{OLED} and field effect transistors\cite{OFET} have been well understood, many open issues are still under debates. Among these, deep and complex issues on the transport of spin polarized carriers or magnetoresistance effects in organic materials have been risen recently by the advent of molecular spintronics\cite{Cinchetti}.

 The transport in organic materials is usually described as a hopping conductivity promoted by polaron hops, and its theoretical description is to some extent similar to the conventional theory of hopping conduction \cite{Efr-Sh}. However this transport is also characterized by a number of features not acknowledged in classical systems with hopping. One of this features is the so-called organic magnetoresistance (OMAR), a strong magnetoresistance easily detectable in relatively weak magnetic fields of $\sim 10-100$Gs. The qualitative understanding of this effect \cite{Kalinowski, Prigodin, Bobbert} is based on the simple observation that the spin relaxation influences the hopping transport, as for example demonstrated by Monte-Carlo simulations\cite{Bobbert}. Nevertheless important basic issues, and especially the role of correlations, remain so far unclear.

The conventional description of the hopping conduction is based on the Miller-Abrahams network. This network and the underlying kinetic equations can be derived from quantum mechanics, while its applicability is limited by several constrains. One of the most important constraint is the so called Hartree decoupling that requires the exclusion of intersite correlations. The average of the product of two filling numbers $\overline{n_in_j}$ is considered to be equal to the product of averaged filling numbers $\overline{n}_i \overline{n}_j$ \cite{Bryksin-Book}. This condition is satisfied in the equilibrium when the Coulomb interaction between charge carriers is not taken into account.

The Coulomb interaction induces nevertheless strong intersite correlations,  leading for example to the phenomenon of the Coulomb glass \cite{Col-glass-1,Col-glass-2,Kogan}. Moreover, the intersite correlations can appear even without the Coulomb interaction when the system is out of the equilibrium. Actually the applied voltage that induces electric current drives the system out of equilibrium and can lead to these correlations. Recently it was shown \cite{Aleiner1,Aleiner2} that the correlations can affect the transport properties of the system even in the linear-response regime.

Here we discuss the correlations in the context of organic magnetoresistance. Specifically we discuss the theory of the bipolaron mechanism of OMAR and show how spin relaxation renormalizes the hopping transport via correlations. The  bipolaron mechanism was first proposed in Ref. \cite{Bobbert} in terms of a smart theoretical model based on the parallel and antiparallel configurations of spins on two sites. The effect was demonstrated with the Monte-Carlo simulations.  This study was followed by several attempts to include these parallel/antiparallel configurations into the conventional theory of hopping conduction \cite{HF1,HF2,HF3,chi-Bassler,Lu2017,Chi1,Larabi}. The number of these attempts itself indicate the interest in generalizing the conventional theory of hopping transport to include the effects similar to OMAR. However all these attempts faced one problem. The conventional approach \cite{Efr-Sh} to the percolation theory follows the scheme: the rate equations are derived first from the quantum mechanics with the Hartree decoupling, then the resistor network is obtained as a linearization of these equations. Only after that the percolation theory is applied to describe the resistivity of this network. However, we showed recently \cite{ourKin} that kinetic equations with Hartree decoupling cannot describe the organic magnetoresistance.  The mentioned studies
\cite{HF1,HF2,HF3,chi-Bassler,Chi1,Larabi} did not provide a re-derivation of the theory. Rather the bipolaron qualitative mechanism of OMAR was artificially included in the conventional percolation picture excluding the intersite correlations. In the present study we show that these correlations naturally include the spin relaxation and are fundamental for the description of the OMAR. Note, that the Monte-Carlo simulations performed in Ref. \cite{Bobbert} automatically include all intersite correlations that are neglected in the Hartree decoupling.

Although we recognize the importance of previous models that fused OMAR with percolation theory we expand the conventional approach by explicitly including the intercite correlations. Namely, we derive the kinetic equations that take into account pair intersite correlations and the possibility of the double occupation. The equations also include spin dynamics in the simplified phenomenological model where it is described by a single spin relaxation time dependent on the magnetic field. We demonstrate that at low applied voltage these equations can be reduced to an effective resistor network. The expression for the effective resistance between sites $i$ and $j$ is more complex than in Miller-Abrahams theory and is dependent on the other sites surrounding the pair $ij$. However this expression depends explicitly on the spin degrees of freedom and can describe the organic magnetoresistance.

The article is organized as follows. In section \ref{sect-OMAR} we give a short qualitative review of the bipolaron mechanism of OMAR. In section \ref{sect-kin} we derive the kinetic equations that include pair intersite correlations in the simple case of large Hubbard energy. We also provide the equations in the general case of arbitrary Hubbard energy. The derivation for the general case is presented in the supplementary materials\cite{supplement}. In the section \ref{sect-res} we linearize the kinetic equations to obtain the effective resistor network. The obtained network generalizes the Miller-Abrahams resistor network by including pair inter-site correlations. It can be used to construct the rigorous percolation theory that describes OMAR. In the section \ref{sect-num} we study the kinetic equations numerically and discuss the main features of OMAR that follow from these equations. In the section \ref{sect-dis} we provide the general discussion of the obtained results.

\section{Bipolaron mechanism of OMAR}
\label{sect-OMAR}

It was proposed \cite{Bobbert} that the organic magnetoresistance can be described with the bipolaron mechanism.
 The main idea of the mechanism is the so called ``spin blocking''. It requires the possibility of double occupation of the hopping site. It means that an electron can hop to a site already occupied with one electron and form a bipolaron provided that the spins of both electrons form a singlet. However, the two electrons with the same spin projections along the common quantization axis have zero singlet probability and cannot form a bipolaron. This leads to
effective reduction of the number of sites available for hopping. The spin relaxation can rotate the electron spins, change the parallel configuration to anti-parallel and therefore restore the possibility of the hop. Note that the magnetic field corresponding to organic magnetoresistance is weak and Zeeman energy is much smaller than temperature. Therefore spin polarization is absent and the number of polarons with up and down spin projections is always the same.  However the spin dynamics still influence the current, because the magnetic field alters the spin relaxation process.

\begin{figure}[htbp]
    \centering
        \includegraphics[width=0.6\textwidth]{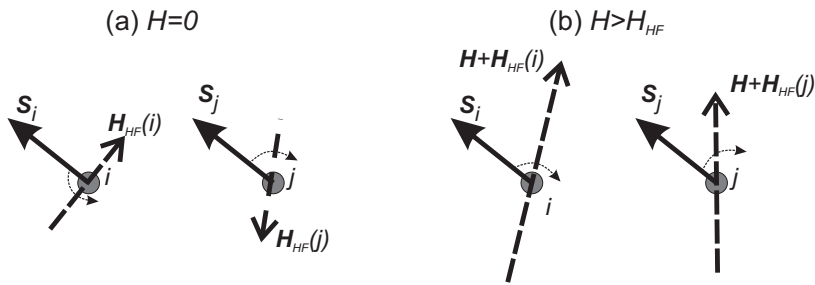}
        \caption{The effect of the applied magnetic field on the hyperfine mechanism of spin relaxation.}
    \label{fig:HF-rel}
\end{figure}

One of the dominant mechanisms of the spin relaxation in organics is the hyperfine interaction with nuclear magnetic moments. The applied magnetic field $\sim 10-100$Gs influences significantly  this mechanism.
The nature of the effect of the applied magnetic field on the spin relaxation rate is shown on Fig. \ref{fig:HF-rel}. The mutual orientation of the spins of electrons on sites $i$ and $j$ relaxes because the spins rotate around the different local hyperfine fields ${\bf H}_{HF}(i)$ and ${\bf H}_{HF}(j)$ (Fig. \ref{fig:HF-rel} (a)). When the applied magnetic field is larger than the hyperfine field (Fig. \ref{fig:HF-rel} (b)) the spins rotate around the total field ${\bf H} + {\bf H}_{HF}$. This total field differs only slightly on sites $i$ and $j$. It significantly suppress the spin relaxation.

The rigorous theoretical description of the mechanism of spin relaxation is quite complex and includes interplay between the polaron hopping and the on-site spin rotation \cite{Movaghar, Yu-HF, HFine2,Harmon-Relax}. It leads to many sophisticated phenomena such as the slow non-exponential tails in the spin relaxation \cite{ourKin, Harmon-Relax}. In the present study we are not specifically interested in the details of the spin relaxation. Rather we focus on the question why the spin relaxation affects the charge transport in the situation when the spin polarization is absent. Therefore we adopt an oversimplified model that reduces the complex physics of the hyperfine mechanism of spin relaxation to the single spin-flip rate $\tau_s^{-1}$ that is dependent on the applied magnetic field.

For this rate we use the expression
\begin{equation}
\tau_s^{-1}=  \omega_s^{(0)}\frac{H_{HF}^2}{H_{HF}^2 + H^2}.
\label{wsrel}
\end{equation}
Here $\omega_s^{(0)}$ is the zero magnetic field spin-flip rate. $H_{HF}$ is the typical value of the hyperfine field and $H$ is the applied magnetic field. The term $H_{HF}^2/(H^2 + H_{HF}^2)$ reflect the suppression of the spin relaxation in the applied magnetic field. Note that the reduction of the hyperfine spin relaxation to the magnetic field-dependent spin-flip rate was used in several studies describing the organic magnetoresistance. The equation (\ref{wsrel}) agrees with the approach used in Refs. \cite{HF1,chi-Bassler}

\section{Kinetic equations with intersite pair correlations}
\label{sect-kin}

Conventionally the derivation of the theory of hopping conduction starts from kinetic equations
\begin{equation} \label{kin1}
\frac{d\overline{n}_{i, \uparrow}}{dt} = \sum_j W_{ji} \overline{n_{j,\uparrow} p_{i,\uparrow}} - W_{ij} \overline{n_{i,\uparrow} p_{j,\uparrow}}.
\end{equation}
Here $\overline{n}_{i, \uparrow}$  is the average filling number of spin-up electron on site $i$. $p_{j,\uparrow}$ stands for the possibility to find an empty place for spin-up electron on site $j$. The mutual line over $n_{i,\uparrow}$ and $p_{j,\uparrow}$ corresponds to the joint averaging of the product $n_{i,\uparrow} p_{j,\uparrow}$.

The next conventional step that impedes the description of the organic magnetoresistance is the Hartree decoupling. This approximation corresponds to the substitution $\overline{n_{i,\uparrow} p_{j,\uparrow}} \rightarrow \overline{n}_{i,\uparrow} \overline{p}_{j,\uparrow}$. It allows to obtain many important physical results, for example the temperature dependence of conductivity and orbital magnetoresistance \cite{Efr-Sh} but is known to be insufficient for the description of OMAR \cite{ourKin}. Here we go one step beyond this approximation. We include pair correlations of the filling numbers but exclude triple correlations. It means that when we consider pair of sites
$i-j$ we make the decoupling $\overline{n_{i\uparrow}n_{j\uparrow}n_{k\uparrow}} \rightarrow \overline{n_{i\uparrow}n_{j\uparrow}} \cdot \overline{n}_{k\uparrow}$. Our assumptions are similar to the approach used recently in Refs. \cite{Aleiner1,Aleiner2}. However, the theory \cite{Aleiner1,Aleiner2} does not include the double occupation possibility that is essential for the bipolaron mechanism of OMAR.

\begin{figure}[htbp]
    \centering
        \includegraphics[width=0.3\textwidth]{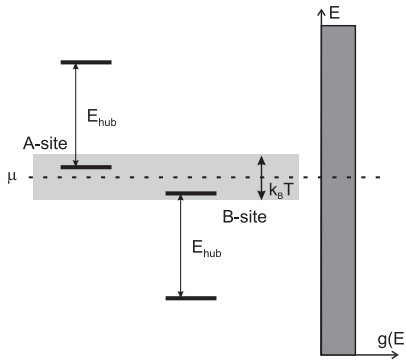}
        \caption{The cartoon for A-type and B-type sites. A-type sites have the single occupation states in the energy band of the width $k_B T$ around the Fermi level (shown with light grey color). B-type sites have the double-occupied states in this band. To provide both types of sites the density of states $g(E)$ should be wider than $E_{hub}$. }
    \label{fig:AB}
\end{figure}

We start our consideration from the simplest model that describe hopping with double occupation possibility. This model assumes that the Hubbard energy $E_{hub}$  is much larger than temperature, but the localization sites have broad energy distribution $g(E)$ with the width larger than $E_{hub}$. In order to effectivly participate in the hopping the localization site should have a state with the energy close to the chemical potential $\mu$. The states far below $\mu$ are always filled and the states with too large energies are never occupied. Only the sites with an energy level in some energy band with the width $\sim k_B T$ around chemical potential can change their filling numbers and effectively participate in hopping. This band is shown by light grey color on Fig. \ref{fig:AB}. More rigorously, in the situation with an exponentially-broad distribution of the hopping rates the width of the band is $\xi_c k_B T$, where $\xi_c$ is the percolation exponent \cite{Efr-Sh}. The condition of large Hubbard energy in this case reads $E_{hub} \gg \xi_c k_B T$ but we limit our analysis to the case $\xi_c \sim 1$.   For these conditions there are two types of sites that are important for conductivity (see Fig. \ref{fig:AB})\cite{matveev}. The so-called A-type sites have their energy $E_i$ near the Fermi energy $E_i \sim \mu$. These sites are never double occupied because $E_i + E_{hub} - \mu \gg k_B T$.   The B-type sites are charactrised by single-occupation energy well below the chemical potential and these sites always have at least one electron. The energy of doubly occupied B-site is near the chemical potential $E_i + E_{hub} \sim \mu$, therefore these sites can have one or two electrons. The exact form of the density of states $g(E)$ is not very important for our study. In figure \ref{fig:AB} we consider $g(E)=const$ in some interval of energies. Nevertheless our consideration is valid for other $g(E)$ that allows the existence of both $A$-type and $B$-type sites. For example it is valid for the model of host and donor states in polymers considered in \cite{bob-polim}.

In our model we consider small magnetic fields that cannot lead to the spin polarization. Therefore the averaged filling numbers of a site for different spin projections are always the same $\overline{n}_{i\uparrow} = \overline{n}_{i\downarrow}$. Also we assume the spin inversion symmetry that leads to the conservation of any averaged value after inversion of all the spin projections, for example $\overline{n_{i\uparrow}n_{j\downarrow}} = \overline{n_{i\downarrow}n_{j\uparrow}}$.  However this symmetry still allows the existence of the spin correlations. For A-site $i$ and B-site $j$ the spin correlation can be defined as $s_{ij} = \overline{n_{i\uparrow} p_{j\uparrow}}-\overline{n_{i\uparrow} p_{j\downarrow}}$. Beside the spin correlations $s_{ij}$  the model allows another kind of pair correlations that are less dependent on the spin degree of freedom. We call them charge correlations $c_{ij}$. For A-B pair of sites $i-j$ it can be defined as $c_{ij} = \overline{n_{i\uparrow} p_{j\uparrow}} + \overline{n_{i\uparrow} p_{j\downarrow}} - 2 \overline{n}_{i\uparrow}\overline{p}_{j\uparrow}$.

\begin{figure}[htbp]
    \centering
        \includegraphics[width=0.3\textwidth]{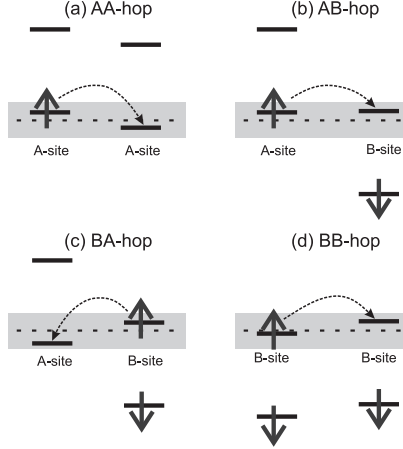}
        \caption{Hopping between different kinds of sites. The AB hop (b) is allowed only for opposite spin directions of the hopping electron and of the electron on ``free'' $B$-type site. Other hops are always possible. The $BA$ hop (c) always results in the antiparallel spin configuration on the sites. }
    \label{fig:ABhop}
\end{figure}

To describe these correlations we should write not only the rate equations for averaged filling numbers $\overline{n}_{i\uparrow}$ but also the equations for their products, for example for $\overline{n_{i\uparrow} p_{j\uparrow}}$. Let us assume that the site $i$ is of type $A$ and site $j$ is of type $B$. In this case $\overline{n_{i\uparrow} p_{j\uparrow}}$ stands for the probability for both sites to have one electron (this situation is shown on the fig. \ref{fig:ABhop} (b)). The electron on site $i$ has spin projection $\uparrow$ and the electron on site $j$ has spin projection $\downarrow$ (it allows the hop of $\uparrow$-electron to the $B$-type site $j$). The configuration $n_{i\uparrow} p_{j\uparrow}$ can appear due to the electron hopping in several different way. First, it can appear due to the hop of spin-up electron from the site $j$ to site $i$ (see fig. \ref{fig:ABhop} (c)). Second, it can appear due to  the hop of spin-up electron from the third site $k$ to site $i$ while the site $j$ already have only one spin-down electron. Third, it can appear due to the hop of spin-up electron from site $j$ to the third site $m$ while the site $i$ is occupied by another spin-up electron. There are also three ways to break the configuration $n_{i\uparrow} p_{j\uparrow}$: the hop $i\rightarrow j$ and the hops of spin-up electron $i\rightarrow k$ or $m \rightarrow j$. Finally the finite spin relaxation time $\tau_s$ leads to another possibility of the appearance and the disappearance of the configuration $n_{i\uparrow} p_{j\uparrow}$ due to the spin flip. Taking into account all the possibilities we write the rate equation for $\overline{n_{i\uparrow} p_{j\uparrow}}$
\begin{equation}\label{nipj}
\frac{d \overline{n_{i\uparrow} p_{j\uparrow}} }{dt} = W_{ji} \overline{p_{i\uparrow} n_{j\uparrow}} + \sum_{k\ne i,j} W_{ki} \overline{n}_{k\uparrow} \overline{p_{i\uparrow} p_{j\uparrow}}  + \sum_{m \ne i,j} W_{jm} \overline{p}_{m\uparrow} \overline{n_{i\uparrow}n_{j\uparrow}} + \frac{1}{\tau_s} \overline{n_{i\uparrow} p_{j\downarrow}} -
\end{equation}
$$
 - \overline{n_{i\uparrow} p_{j\uparrow}} \left( \frac{1}{\tau_s} +  W_{ij} + \sum_{k\ne i,j} W_{ik} \overline{p}_{k\uparrow} + \sum_{m\ne i,j} W_{mj} \overline{n}_{m\uparrow} \right).
$$
In this equation we took into account that $\overline{n_{i\uparrow} p_{j\downarrow}} = \overline{n_{i\downarrow} p_{j\uparrow}}$ due to the spin inversion symmetry. The values ${n}_{i\uparrow}$ and $p_{i\downarrow}$ are of course not independent. However their relation is different for $A$-type and $B$-type sites. For $A$-type site $i$, $p_{i\uparrow}$ corresponds to the situation when the site $i$ has no electrons. Therefore we can substitute $p_{i\uparrow} = p_{i\downarrow} = 1-n_{i\uparrow}-n_{i\downarrow}$. We can make this substitution under the joint averaging or when all the terms are averaged separately. On the other hand the substitution $n_{i\uparrow} = n_{i\downarrow}$ is possible only  for separate averaging but not under the joint averaging. It means that $\overline{n}_{i\uparrow} = \overline{n}_{i\downarrow}$ but $\overline{n_{i\uparrow} p_{j\uparrow}} \ne \overline{n_{i\downarrow} p_{j\uparrow}}$.  For the $B$-type site $j$ the value $n_{j\uparrow}$ correspond to the situation when the site $j$ is double-occupied and the substitution can be made $n_{j\uparrow} = n_{j\downarrow} = 1 - p_{j\uparrow} - p_{j\downarrow}$ both under the joint or the separate averaging of the terms. The substitution $p_{j\uparrow} \rightarrow p_{j\downarrow}$ is not possible under the joint averaging.

In the expression (\ref{nipj}) we used the decoupling $\overline{n_{i\uparrow}n_{j\uparrow} p_{m\uparrow}} \rightarrow \overline{n_{i\uparrow}n_{j\uparrow}} \, \overline{p}_{m\uparrow}$. Strictly speaking the correct decoupling taking into account all the pair correlations and neglecting triple correlations is $\overline{n_{i\uparrow}n_{j\uparrow} p_{m\uparrow}} \rightarrow \overline{n_{i\uparrow}n_{j\uparrow}}\, \overline{p}_{m\uparrow} + \overline{n}_{i\uparrow}\overline{n_{j\uparrow} p_{m\uparrow}} + \overline{n}_{j\uparrow}\overline{n_{i\uparrow} p_{m\uparrow}} - 2 \overline{n}_{j\uparrow}\overline{n}_{i\uparrow} \overline{p}_{m\uparrow}$. This substitution becomes more clear if we express the averaged product $\overline{n_{i\uparrow}n_{j\uparrow} p_{m\uparrow}}$ in terms of product of averaged filling numbers $\overline{n}_{j\uparrow}\overline{n}_{i\uparrow} \overline{p}_{m\uparrow}$, double and triple correlations and neglect the triple correlation. Here we neglect "long range" correlations. If site $m$ is connected by the hopping with site $j$ we consider the hopping $i-m$ to be irrelevant for the dynamics of correlations $i-j$. In this case we neglect the correlations $i-m$. This assumption is similar to the one made in \cite{Aleiner1,Aleiner2}.  With this assumption when we go from the averaged joint products to the correlations $s_{ij}$ and $c_{ij}$ the additional terms corresponding to the pair correlations of sites $i$ and $j$ with other sites will be canceled. Therefore we do not write them from the beginning.

We believe that it is instructive to exclude $p_{i}$ and $n_j$ from the expression (\ref{nipj}) to keep only the electron notations for the $A$-type site $i$ and hole notations for $B$-type site $j$. However we will keep both $p_k$ and $n_k$ for sites $k$ and $m$ because these sites play only auxiliary role in the determination of the correlations $s_{ij}$  and $c_{ij}$ and we are not particularly interested in their type.
\begin{equation}\label{nipj2}
\frac{d \overline{n_{i\uparrow} p_{j\uparrow}} }{dt} =  \frac{1}{\tau_s} \overline{n_{i\uparrow} p_{j\downarrow}} +
W_{ji}(1-2\overline{n}_{i\uparrow} - 2\overline{p}_{j\uparrow} +  2\overline{n_{i\uparrow} p_{j\uparrow}} + 2\overline{n_{i\uparrow} p_{j\downarrow}}) + \sum_{k \ne i,j} W_{ki}\overline{n}_{k\uparrow} (\overline{p}_{j\uparrow} - \overline{n_{i\uparrow}p_{j\uparrow}} - \overline{n_{i\uparrow}p_{j\downarrow}}) +
\end{equation}
$$
+ \sum_{m \ne i,j} W_{jm} \overline{p}_{m\uparrow} (\overline{n}_{i\uparrow} - \overline{n_{i\uparrow}p_{j\uparrow}} - \overline{n_{i\uparrow}p_{j\downarrow}}) - \overline{n_{i\uparrow} p_{j\uparrow}} \left( \frac{1}{\tau_s} +  W_{ij} + \sum_{k\ne i,j} W_{ik} \overline{p}_{k\uparrow} + \sum_{m\ne i,j} W_{mj} \overline{n}_{m\uparrow} \right).
$$

To get the equations for correlations $s_{ij}$ and $c_{ij}$ it is necessary to write a similar equation for $\overline{n_{i\uparrow} p_{j\downarrow}}$ and combine it with equation (\ref{nipj2}) and with the rate equations for averaged filling numbers $\overline{n}_{i\uparrow}$ and $\overline{p}_{j\uparrow}$. The straightforward algebra yields
\begin{equation}\label{sij}
\frac{d s_{ij}}{dt} = J_{ji} - r_{s,ij}s_{ij}, \quad r_{s,ij} = \frac{1}{\tau_s} + r_{s,ij}^{(0)}, \quad r_{s,ij}^{(0)} =  \sum_{k \ne i,j} W_{ik} \overline{p}_{k\uparrow} + \sum_{m \ne i,j} W_{mj}\overline{n}_{m\uparrow},
\end{equation}
\begin{equation}\label{cij}
\frac{d c_{ij}}{dt} = J_{ji}(1 - 2\overline{n}_{i\uparrow} - 2\overline{p}_{j\uparrow}) - r_{c,ij}c_{ij},
\end{equation}
\begin{equation} \label{Rc}
\quad r_{c,ij} = \sum_{k \ne i,j} \left(2 W_{ki} \overline{n}_{k\uparrow} +  W_{ik} \overline{p}_{k\uparrow} \right) + \sum_{m \ne i,j} \left( 2W_{jm} \overline{p}_{m\uparrow} + W_{mj}\overline{n}_{m\uparrow}\right).
\end{equation}
In this equations $J_{ji} = W_{ji}(1-2\overline{n}_{i\uparrow} - 2\overline{p}_{j\uparrow} +  2\overline{n_{i\uparrow} p_{j\uparrow}} + 2\overline{n_{i\uparrow} p_{j\downarrow}})  - W_{ij}\overline{n_{i\uparrow} p_{j\uparrow} }$ is the current between sites $j$ and $i$ carried by spin-up electrons. Due to spin inversion symmetry this current is exactly equal to the current carried by spin-down electrons. Therefore the total current $j \rightarrow i$ is equal to $2J_{ji}$. The current acts as a source that generates correlations $s_{ij}$ and $c_{ij}$. The correlations relax with the rates $r_{s,ij}$ and $r_{c,ij}$ respectively. The relaxation is due to the hops between sites $i$,$j$ and other sites of the system. The relaxation rate of the spin correlation $r_{s,ij}$ contains the term $1/\tau_s$ that is related to the spin relaxation time and is dependent on the applied magnetic field.

To make the system of equations complete we should write the expression for $J_{ji}$ in terms of correlations $s_{ij}$ and $c_{ij}$.
\begin{equation} \label{Jji}
J_{ji} = W_{ji}(1-2\overline{n}_{i\uparrow} - 2\overline{p}_{j\uparrow} +  4\overline{n}_{i\uparrow} \overline{p}_{j\uparrow} + 2 c_{ij}) - W_{ij} \left( \overline{n}_{i\uparrow} \overline{p}_{j\uparrow} + \frac{c_{ij} + s_{ij}}{2} \right).
\end{equation}

These equations give the general picture of the effect of correlations on the transport properties. The correlations $i-j$ are absent in equilibrium state. They are generated by current $J_{ji}$ as it is shown in Eqs. (\ref{sij},\ref{cij}) and have some relaxation rate related to the hopping between sites $i$, $j$ and other sites of the system. The correlations $s_{ij}$ and $c_{ij}$ enter the expression (\ref{Jji}) for the current indicating that the correlations can alter the transport properties of the system. Note that the correlations enter Eq.(\ref{Jji}) in addition to the terms with averaged filling number. It means that in the non-equilibrium state of the system where all the filling numbers are equilibrium but the correlations $c_{ij}$ and $s_{ij}$ are non-zero, the current can flow in the system. It is important to underline that the correlations cannot be neglected even in linear response regime. In \cite{ourKin} it was presumed that the correlations should influence the effective "resistances" and the transport only at high voltages.

The relaxation rate of the correlation $s_{ij}$ is dependent on $\tau_s$ and therefore the spin relaxation time influence the transport properties even when the magnetization is zero, i.e. in the situation of the spin inversion symmetry.

These expressions correspond to the situation when the site $i$ has type $A$ and the site $j$ has type $B$. In the opposite situation when the site $i$ has type $B$ and site $j$ has type $A$ the equations (\ref{sij} -\ref{Jji}) can still be used with the substitution $i \leftrightarrow j$. For example the spin correlation $s_{ij}$ will be generated not with the current $J_{ji}$  but with current $J_{ij} = -J_{ji}$. In the situation when the two sites have the same type the expressions for the correlations are different.

Let us consider the two $A$-type sites $i$ and $i'$ (fig. \ref{fig:ABhop} (a)). The two correlations of the filling numbers on these sites can be defined as $s_{ii'} = \overline{n_{i\uparrow}n_{i'\uparrow}} - \overline{n_{i\uparrow}n_{i'\downarrow}}$ and
$c_{ii'} = \overline{n_{i\uparrow}n_{i'\uparrow}} + \overline{n_{i\uparrow}n_{i'\downarrow}} - 2\overline{n}_{i\uparrow}\overline{n}_{i'\uparrow}$. It appears that the correlation $s_{ii'}$ has no source, i.e. its expression is $d s_{ii'}/dt = -r_{s,ii'} s_{ii'}$. Therefore although this correlation can in principle exist  in a non-equilibrium state, it is not generated by the applied voltage. We will consider that such correlations are absent in the system. The correlation $c_{ii'}$ is generated by the current and can influence the current.
\begin{equation} \label{c_AA}
\frac{d c_{ii'}}{dt} = 2(\overline{n}_{i\uparrow} - \overline{n}_{i'\uparrow})J_{i'i} - r_{c,ii'}c_{ii'}, \quad
r_{c,ii'} = \sum_{k\ne i,i'} \left( 2W_{ki} \overline{n}_{k\uparrow} + W_{ik}\overline{p}_{k\uparrow} \right) +
\sum_{m\ne i,i'} \left( 2W_{mi'} \overline{n}_{m\uparrow} + W_{i'm}\overline{p}_{m\uparrow} \right),
\end{equation}
\begin{equation} \label{J_AA}
J_{i'i} = W_{i'i}(\overline{n}_{i'\uparrow} - 2\overline{n}_{i'\uparrow} \overline{n}_{i\uparrow} - c_{ii'}) -
W_{ii'}(\overline{n}_{i\uparrow} - 2 \overline{n}_{i\uparrow} \overline{n}_{i'\uparrow} - c_{ii'}).
\end{equation}

The situation in pairs of $B$-sites is similar to the situation in pairs of $A$-sites. In the pair $j-j'$ of two $B$-type sites (fig. \ref{fig:ABhop} (d)) one can consider two correlations $s_{jj'} = \overline{p_{j\uparrow}p_{j'\uparrow}} - \overline{p_{j\uparrow}p_{j'\downarrow}}$ and
$c_{jj'} = \overline{p_{j\uparrow}p_{j'\uparrow}} + \overline{p_{j\uparrow}p_{j'\downarrow}} - 2\overline{p}_{j\uparrow}\overline{p}_{j'\uparrow}$. However the correlation $s_{jj'}$ is not generated by the current and does not contribute to the transport properties. The correlation $c_{jj'}$ is generated by the current and can influence the transport properties
\begin{equation} \label{c_BB}
\frac{d c_{jj'}}{dt} = 2(\overline{p}_{j'\uparrow} - \overline{p}_{j\uparrow})J_{j'j} - r_{c,jj'}c_{jj'}, \quad r_{c,jj'} =
\sum_{k\ne j,j'} \left( W_{kj} \overline{n}_{k\uparrow} + 2W_{jk}\overline{p}_{k\uparrow} \right) +
\sum_{m\ne j,j'} \left( W_{mj'} \overline{n}_{m\uparrow} + 2W_{j'm}\overline{p}_{m\uparrow} \right),
\end{equation}
\begin{equation} \label{J_BB}
J_{j'j} =W_{j'j}(\overline{p}_{j\uparrow} - 2 \overline{p}_{j\uparrow}\overline{p}_{j'\uparrow} - c_{jj'})
- W_{jj'}(\overline{p}_{j'\uparrow} - 2 \overline{p}_{j\uparrow}\overline{p}_{j'\uparrow} - c_{jj'}).
\end{equation}

 The equations (\ref{sij}-\ref{J_BB}) are the kinetic equations for the case of infinite Hubbard energy (the model of A and B-type sites). In this case each pair of sites $i$ and $j$ can be described with four values: $\overline{n}_{i\uparrow}$, $\overline{n}_{j\uparrow}$,  the spin correlation $s_{ij}$ and the charge correlation $c_{ij}$. The similar considerations is possible also for the general case of arbitrary Hubbard energy.
In this case each site can have zero, one or two electrons and can be described by two independent filling numbers $\overline{n}_{i\uparrow}$ and $\overline{n}_{i2}$ --- the probability for site $i$ to have one electron with spin up and two electrons respectively. The number of independent correlations $\nu_{ij}$ between sites $i$ and $j$ is equal to five. We describe them with the vector $\vec{\nu}_{ij}$
\begin{equation}\label{nu-cor} \vec{\nu}_{ij} = \left( \begin{array}{c} s_{ij}\\ c_{ij}\\ \nu_{ij,2\uparrow}\\ \nu_{ij,\uparrow2}\\ \nu_{ij,22} \end{array}
\right) =
\left(
\begin{array}{c}
s_{ij}\\ c_{ij}\\
\overline{n_{i2} n_{j\uparrow}}  -
\overline{n}_{i2}\overline{n}_{j\uparrow} \\ \overline{n_{i\uparrow} n_{j2}}  - \overline{n}_{i\uparrow}\overline{n}_{j2} \\ \overline{n_{i2} n_{j2}}  - \overline{n}_{i2}\overline{n}_{j2}
\end{array}
\right)
\end{equation}
The additional correlations are related to the possibility for each site to play both roles: of $A$-site and of $B$-site.
 These additional correlations become proportional to $c_{ij}$ in the limit $E_{hub} \rightarrow \infty$. Only one correlation $s_{ij}$ is directly related to the spin relaxation rate. Therefore the new correlations can be considered as additional charge correlations.

In this general case it is useful to introduce four currents flowing between sites $i$ and $j$,  $J_{ij}^{X}$ where indexes $X$ describe the role played by the sites and can have one of four values $AA$, $AB$, $BA$ or $BB$. For example the current $J_{ij}^{AB}$ stands for the hop of the first electron on site $i$ to the single-occupied site $j$ and to the backward hop of second electron from site $j$ to the empty site $i$. All four currents are equal to zero in the equilibrium.

The general picture of correlations affecting the electron kinetics is similar to the one in the model of $A$-type and $B$-type sites. The correlations $\nu_{ij}^{\alpha}$ are generated by the currents $J_{ji}^{X}$ and relax due to hops to other sites and the spin relaxation.
\begin{equation}\label{cor-mat1}
\frac{d}{dt} \nu_{ij}^{\alpha} = G_{ij}^{\alpha X} J_{ji}^{X} - R_{ij}^{\alpha\beta}\nu_{ij}^{\beta}
\end{equation}
Here the indexes $\alpha$ and $\beta$ stand for the components of vector $\vec{\nu}_{ij}$.
The correlations $\vec{\nu}_{ij}$ contribute to the currents in the additive way \begin{equation}\label{cor-cur1} J_{ji}^X = J_{ji,0}^X + {\cal W}_{ij}^{X\alpha} \nu_{ij}^{\alpha} \end{equation} Here $J_{ji,0}^X$ are the expression for currents that neglect correlations.
$G_{ij}^{\alpha X}$, $R_{ij}^{\alpha\beta}$ and ${\cal W}_{ij}^{X\alpha}$ are the matrixes that describe the generation of correlations by currents, relaxation of correlations and the effect of correlations on currents correspondingly. The explicit form of these matrixes is rather cumbersome and we present them in suplemental materials\cite{supplement} along with the explicit expression for $J_{ji,0}^X$.

\section{Resistor network}
\label{sect-res}

The kinetic equation for the hopping transport can be linearized in the limit $eEr_{ij} \ll k_B T$ where $E$ is the applied electric field and $r_{ij}$ is the distance between hopping sites. In the conventional hopping theory this linearization yields the Miller-Abrahams resistor network. In the present section we show that the linearization of our equations (\ref{sij}-\ref{J_BB}) for the model of A and B-type sites leads to the generalized resistor network where the conductivities of resistors connecting $A$ and $B$ sites explicitly depend on the spin relaxation time $\tau_s$. Let us note that  the resistor network approach was applied for the description of OMAR in \cite{bob-polim} where a node of the network was related to the many-body states of the system. This approach is different from our generalized Miller-Abrahams network where the nodes of the network correspond to the localization sites.

In the linearized equation the relaxation rates $r_{s,ij}$ and $r_{c,ij}$ should be calculated in the equilibrium. Therefore the equilibrium values of $\overline{n}_{k\uparrow}$ and $\overline{p}_{k\uparrow}$ should be substituted in (\ref{sij}) and (\ref{Rc}). The relaxation rates $r_{s,ij}$ and $r_{c,ij}$ then appear to have constant values determined by the configuration of the disorder.

The equations (\ref{sij}) and (\ref{cij}) can be reduced to a matrix equation for the correlators
\begin{equation}\label{sc-matrix}
\left(
\begin{array}{cc}
r_{s,ij} + \frac{W_{ij}}{2} & \frac{W_{ij}}{2} - 2W_{ji} \\
\frac{W_{ij}}{2} & \widetilde{r}_{c,ij} + \frac{W_{ij}}{2} - 2W_{ji}
\end{array}
\right)
\left(
\begin{array}{c}
s_{ij} \\ c_{ij}
\end{array}
\right)
=
\left(
\begin{array}{c}
1 \\ 1
\end{array}
\right)
J_{ji}^{(0)}
\end{equation}
$$\widetilde{r}_{c,ij} = r_{c,ij}/(1-2\overline{n}_{i\uparrow} - 2\overline{p}_{j\uparrow})$$
Here $J_{ji}^{(0)}$ is the term in the current that does not include the correlators. It can be expressed in terms of Miller-Abrahams resistance $R_{ij}^{(MA)}$ of the pair $i-j$ and the voltage $u_{ji}$ applied to the pair
$J_{ji}^{(0)} = u_{ji}/R_{ij}^{(MA)}$. $\widetilde{r}_{c,ij} = r_{c,ij}/(1-2\overline{n}_{i\uparrow} - 2\overline{p}_{j\uparrow})$.

The equation (\ref{sc-matrix}) should be solved and the values of the correlators $c_{ij}$ and $s_{ij}$ should be substituted to the equation (\ref{Jji}). It leads to the following equations for resistor that includes the effect of pair inter-site correlations
\begin{equation}\label{RAB}
R_{ij}^{(AB)} = R_{ij}^{(MA)} \left( 1 + \frac{W_{ij}}{(2/\tau_s) + 2r_{s,ij}^{(0)}} + \frac{W_{ij} - 4W_{ji}}{2 \widetilde{r}_{c,ij}} \right).
\end{equation}
The equation (\ref{RAB}) demonstrates that the correlations enter the expression for the resistor as an additional multiplier. This multiplier contains the term $W_{ij}/(2/\tau_s + 2r_{s,ij}^{(0)}) = W_{ij}/2{r}_{s,ij}$ that depends on the applied magnetic field. It is also clear from the equation (\ref{RAB}) that the discussed mechanism of OMAR leads to the positive magnetoresistance. Naturally, $r_{s,ij}$ decreases with applied magnetic field leading to the increase in the resistance $R_{ij}^{(AB)}$ because the term $W_{ij}/2{r}_{s,ij}$ is always positive. The second term $(W_{ij} - 4W_{ji})/2\widetilde{r}_{c,ij}$ is related to charge correlations $c_{ij}$. It can have arbitrary sign but does not depend on $\tau_s$ and on the applied magnetic field.

The resistors that connect pairs of sites of equal type can be treated in the same way. The corresponding expression for resistances $R_{ii'}^{(AA)}$ and $R_{jj'}^{(BB)}$ are
\begin{equation}\label{RAA}
R_{ii'}^{(AA)} = R_{ii'}^{(MA)} \left(1 + \frac{2(\overline{n}_{i\uparrow} - \overline{n}_{i'\uparrow})(W_{i'i} - W_{ii'})}{r_{c,ii'}} \right).
\end{equation}
\begin{equation}\label{RBB}
R_{jj'}^{(BB)} = R_{jj'}^{(MA)} \left(1 + \frac{2(\overline{p}_{j\uparrow} - \overline{p}_{j'\uparrow})(W_{jj'} - W_{j'j})}{r_{c,jj'}} \right).
\end{equation}
The resistances $R_{ii'}^{(AA)}$ and $R_{jj'}^{(BB)}$ are dependent on the charge correlation $c_{ii'}$ but not on spin correlation $s_{ii'}$. Therefore magnetic field does not enter the expression for these resistances.

The equations (\ref{RAB}-\ref{RBB}) reduce the problem of A and B-type sites at low applied voltage to the network of classical resistors. This network can be treated by the same method as a classical Miller-Abrahams resistor network, for example with percolation theory or with direct numerical solution of the Kirchhoff equations that is much easier than the Monte-Carlo simulation.

The linearization of the kinetic equations in the case of arbitrary Hubbard energy is discussed in supplemental materials \cite{supplement}. In the general case the linearization yields the analog of the Kirchhoff equations --- the system of linear equations that can be solved to find the site potentials and the currents. However the linear equations in this case cannot be reduced to an equivalent scheme that contains only resistances.

\section{Numerical simulation}
\label{sect-num}

The kinetic equations Eqs.(\ref{cor-mat1}-\ref{cor-cur1}) describe the microscopic responce of pairs of sites to the applied electric field. The calculation of the magnetoresistance in a macroscopic sample requires an averaging over the sample. In principle the linearized version of kinetic equation allows to apply conventional methods of averaging such as the percolation theory in a resistance network with exponentially-broad distribution of resistances \cite{Efr-Sh}, but this is beyond the scope of the present study.

We follow an alternative way and solve the kinetic equations numerically. We consider the general case, i.e. do not restrict our simulations to the model of $A$ and $B$-type sites and to small applied electric fields. The main goal of the present simulation is to show the most general features of the magnetoresistance described with the kinetic equations (\ref{cor-mat1}-\ref{cor-cur1}) and compare them with the other existing models, such as the percolation theory based on the momentary filling numbers \cite{HF1,HF2,HF3}.

\begin{figure}[htbp]
    \centering
        \includegraphics[width=0.3\textwidth]{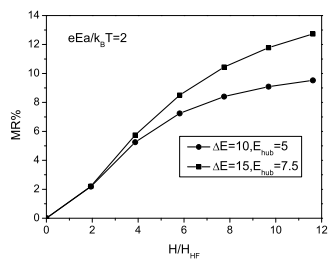}
        \caption{The dependence of the magnetoresistance on magnetic field. Here we assume that the spin relaxation rate is determined by equation  (\ref{wsrel}) with $\omega_{s}^2/\omega_0^2=1/15$.}
    \label{fig:MR}
\end{figure}

In our simulations we apply standard Euler method. Similar to Ref.\cite{bobbert2014} we apply
periodic boundary conditions in the presence of the external electric field. For numerical
simulations we use a square lattice with the size $32\times32$ and $64\times64$ sites with energetic disorder $-\Delta E/2 \le E_{i,j} \le \Delta E/2$ and with
Miller-Abrahams nearest neighbors hopping rates
$W_{i,j}= \omega_0\exp{(-[\Delta E_{i,j}+|\Delta E_{i,j}|]/2k_BT)}$,  where $\omega_0$ is a prefactor and $\Delta E_{i,j}$ is the energy difference between sites $j$ and $i$ including the contribution from the external electric field and Hubbard energy $E_{hub}$ in the case of double occupation. The logarithms of the conductances \cite{bobbert2014} are averaged over 100 random energy configurations. To make sure that the size effects are small we have compared the calculated magnetoresistance for the systems $32\times32$ and $64\times64$ sites. As it follows from the comparison presented in suplemental materials\cite{supplement} the finite size effects are negligible. Similarly we have compared the results for magnetoresistance calculated for different number of averaging over random energy configurations. The results presented in the supplemental materials\cite{supplement} suggests that averaging over 100 random energy configurations provides reliable results for the considered parameters of the system.
The results of calculations of the magnetoresistance are presented in Fig.\ref{fig:MR}.
The magnetoresistance quickly increases at small magnetic field $H\le H_{HF}$and then slowly approaches
its limiting value $MR\approx MR_{\infty}-const  H^{-2}$ at $H\gg H_{HF}$. Note that in the presented case the limiting magnetoresistance is about $MR_{\infty}\approx 10\%$ but it is increasing with the increase of disorder and Hubbard energy.

It is interesting that the magnetoresistance is absent for the absence of disorder $\Delta E=0$ and zero Hubbard energy $E_{hub}=0$. It can be shown analytically that in this case the equilibrium filling numbers and zero inter-site correlations solve the kinetic equations in arbitrary applied field. Therefore no organic magnetoresistance can be observed in such a system. Indeed, assuming that $n_{i2}$ and $n_{i\uparrow}$ are equilibrium filling numbers and taking into account that $n_{i2}(1-2n_{i\uparrow}-n_{i2})=n_{i\uparrow}^2$ we can write the expression for uniform currents: $J_{ij}^{AA}=\omega_0n_{i\uparrow}(1-2n_{i\uparrow}-n_{i2})(1-\exp{(-eEa/k_BT)})$, $J_{ij}^{AB}=J_{ij}^{BA}=\omega_0n_{i\uparrow}^2(1-\exp{(-eEa/k_BT)})$, $J_{ij}^{BB}=\omega_0n_{i\uparrow}n_{i2}(1-\exp{(-eEa/k_BT)})$. Substituting these currents to Eq.(\ref{cor-mat1}) it is easy to see that the nonuniform part of this equation containing currents reduces to zero. It means that all the correlations go to zero $\nu_{i,j}^{\alpha}=0$. On the other hand the charge conservation equations (Eqs. (22,23) from the supplementary materials \cite{supplement}) are satisfied automatically for uniform currents.
Physically it means that although the hopping rates are modified by the applied electric field and the current flows, the distribution of charges and spins remains equilibrium. The spin relaxation has no effect on the equilibrium distribution of spins and does not influence the charge transport. Note, however, that this is the case only for relatively simple ordered systems like the ordered square lattice. In the system with two or more types of site with different energies forming some sort of complex lattice the organic magnetoresistance should be nonzero.
To the best of our knowledge the absence of OMAR in these conditions was never reported and it was not predicted by models \cite{HF1,HF2,HF3}.

\begin{figure}[htbp]
    \centering
        \includegraphics[width=0.3\textwidth]{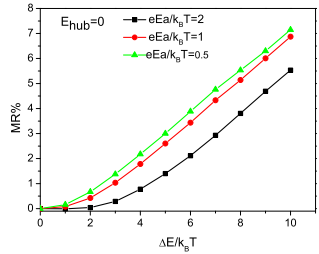}
        \caption{The dependence of the magnetoresistance on disorder for different electric fields.}
    \label{fig:MRdE}
\end{figure}

\begin{figure}[htbp]
    \centering
        \includegraphics[width=0.3\textwidth]{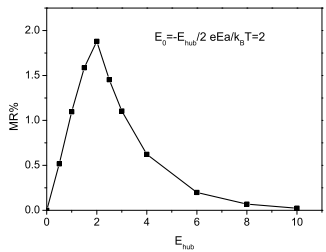}
        \caption{The dependence of the magnetoresistance on the Hubbard energy. All the sites are considered to have the same single-occupation energy $E_0 = -E_{hub}/2$. }
    \label{fig:MRotEhub}
\end{figure}

The increase of disorder in the absence of $E_{hub}=0$ leads to the increase of magnetoresistance
Fig.\ref{fig:MRdE}.
Similarly magnetoresistance appears when the disorder is absent $\Delta E=0$ and $E_{hub}$
is finite Fig.\ref{fig:MRotEhub}. Note that the magnetoresistance has characteristic maximum
when $E_{hub}\approx eEa/k_B T$. Further increase of Hubbard energy leads to the decrease of magnetoresistance.
These results reflect the absence of correlations in ordered system with zero
Hubbard energy. The increase of the disorder and the finite Hubbard energy lead to the appearance of
nonzero correlations and to finite magnetoresistance.

\begin{figure}[htbp]
    \centering
        \includegraphics[width=0.3\textwidth]{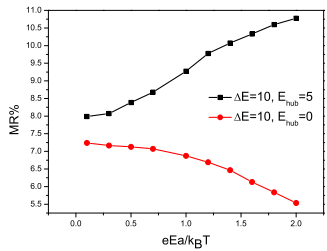}
        \caption{The dependence of the magnetoresistance on external electric field.}
    \label{fig:MRotE}
\end{figure}

Predicted magnetoresistance is finite in the limit of the weak field $eEa/k_BT\ll1$ and has relatively weak dependence on the external electric field.
It increases by 30\% in the  limit $eEa/k_BT\gg1$ for $E_{hub}>eEa$  and decreases by 30\% for the case $E_{hub}<eEa$ Fig. \ref{fig:MRotE}. The increase of the magnetoresistance in the limit of finite $E_{hub}$ with increasing electric field agrees with results obtained in Ref.\cite{Bobbert}.

\section{Discussion}
\label{sect-dis}

Some of the previous attempts to include OMAR into the analytical theory of the hopping transport considered the percolation theory with momentary spin projections that effectively corresponds to the percolation with momentary filling numbers. For example the number of sites available for the hop was counted as the number of free sites and the sites with one electron with the spin antiparallel to the spin of the hopping electron. While such a scheme leads to the finite magnetoresistance it is inconsistent with the general rules that are used in the well-known problem of hopping transport. To clarify that let us consider the simple problem of neighbor hopping over sites with random positions without double occupation possibility and assume that the number of electrons is half the number of sites. The idea of the percolation theory with the momentary filling numbers means that only half of sites are available for the hop because other half is filled. However in the percolation theory for this problem \cite{Efr-Sh} all the sites are counted because a filled site close to the initial site of the hop will typically lose its electron faster than the time of the hop to a more distant site (that is exponentially larger than the time of the hop to the neighbor site). The previous approaches \cite{HF1,HF2,HF3}  do not consider this possibility for the target site of the hop (for example with one electron with the parallel spin configuration) to lose its electron due to some other hop. Note that previous attempts to make an analytical theory of bipolaron mechanism of OMAR did not link this magnetoresistance with the intersite correlations of the filling numbers. Here we show that the relation is quite direct. The absence of correlations, for example in the ordered system with zero Hubbard energy, means the absence of magnetoresistance.

We  propose another approach to include OMAR into analytic theory of hopping transport via including the correlations in the kinetic equations. The shape of the dependence of resistivity on the applied magnetic field in our theory is similar to the one in previous theories. It is governed by the dependence $\tau_s(H)$ that is not closely related to the physics of hopping conduction. However the dependence of the magnetoresistance on system parameters is different. Most clear example of this difference is the absence of the magnetoresistance in the ordered system with zero Hubbard energy.

In our theory we considered only pair intersite correlations neglecting the triple correlations. We also neglected "long range" correlations between sites that are not "connected" with effective hopping. The main reason for these assumptions is that kinetic equations with pair correlations is the minimal model that allows us to include OMAR into conventional theory of the hopping transport. The neglected correlations can in principle modify the organic magnetoresistance but cannot completely suppress it. Note that the neglecting of the triple correlations is a controllable approximation when the number of electrons is small and the number of B-type sites is small (in the model of A-type and B-type sites). In this case the problem allows the expansion into series over the correlation order. Organic magnetoresistance appears in the second order of the expansion (pair correlations). The higher order correlations can lead only to small perturbation in this situation.

Another assumption made in our theory is the simplified treatment of the spin dynamics that was reduced to single spin relaxation time. To consider the effect of realistic spin rotation in the hyperfine field on the conductivity one should include this rotation into kinetic equations for the correlations. However it will not only influence the dynamics of the discussed spin correlation $s_{ij}$ but also will lead to the appearance of new spin correlations. It is similar to the appearance of new charge correlations due to finite Hubbard energy. In general case the spin state of two electrons can be described by $4\times 4$ density matrix. However in our case the system is invariant under   rotations of the spin space. It significantly reduce the possible form of the density matrix. The only non-equilibrium form of the matrix in stationary case is related to the possible imbalance between triplet and singlet electron pairs. It can be expressed in terms of a single spin correlation. However if we introduce real hyperfine fields on the localization sites, these fields will introduced a preferred spin direction. New spin correlations that reflect the appearance of the average values similar to $\langle[({\bf S}_i + {\bf S}_j)({\bf H}_i - {\bf H}_j)]^2\rangle$ are important for the problem. Therefore to make a correlation-based theory of OMAR one should identify the spin correlations that appear in general case and discuss their dynamic in the presence of the local hyperfine fields, the spin-orbit coupling and the exchange interaction.

The absence of the bipolaron OMAR in an ordered system with zero Hubbard energy is not sensitive to the neglecting of the higher-order correlations. Naturally the correlations of different order can be connected with the Bogoliubov-Born-Green-Kirkwood-Yvon chain of equations\cite{Balescu}. This chain of equations shows that the correlations of the order $n+1$ can be generated by the correlations of the order $n$. We have shown that pair correlations are not generated in the ordered system with $E_{hub}=0$. Triple correlations will not be generated because of the absence of double correlations to generate them. The same applies to the realistic spin dynamics. This dynamics should not influence the density matrix that describes the equilibrium distribution of spins and charges. Before the dynamics of the spin correlations becomes important, these correlations should be generated. Note that the only spin correlations that can be generated due to the hopping is the difference between the singlet and triplet probabilities that is proportional to $s_{ij}.$ In the ordered system with $E_{hub} = 0$ the spin correlations are not generated and their spin dynamics is not essential.

The bipolaron mechanism discussed in the present study is the most probable mechanism of the organic magnetoresistance in materials with single carrier type. However in many experiments OMAR was measured in bipolar
devices where the current is carried by both electrons and holes. The physics of such devices is more complex. The effect of the magnetic field on the current can be related to the formation and dissociation of excitons \cite{Kalinowski,Prigodin,AdvMat} or to the interaction of excitons with charge carriers \cite{HUWU,Desai}. In many cases it is still attributed to the effect of magnetic field on the rate of the hyperfine spin relaxation. The difference from the bipolaron mechanism is related to the fact that the mutual spin orientation of the charge carriers influences not the process of bipolaron formation but other processes such as the formation of excitons from  electron-hole pairs. We believe that the method proposed in the present study can be generalized to account for this mechanism. However in this case both electron and hole filling numbers should be included into the kinetic equations along with the corresponding correlations between filling numbers of electrons and holes.

In conclusion, we derive the kinetic equations for the hopping conduction with the double occupation possibility that include pair correlation. Contrary to the conventional equations that are based on Hartree decoupling our equations reflect the dependence of the conductivity on spin relaxation time even without net polarization and can describe OMAR. In the linear regime the equations can be reduced to the generalized resistor network. OMAR described with our equation have a dependence on the magnetic field similar to the previous studies but have different dependence on system parameters.

A.V.Shumilin acknowledges partial support from RFBR (project 16-02-00064 A).

\vspace{3cm}

\hspace{6cm} \textbf{Supplemental materials}
\vspace{1cm}

\hspace{5cm} \textbf{Kinetic equations for arbitrary Hubbard energy.}
\vspace{2cm}

If the Hubbard energy is not large, any site can be in one of the four states (with and without spin up and spin down electrons). To describe this situation we introduce three filling numbers $n_{i\uparrow}$, $n_{i\downarrow}$ and $n_{i2}$.  $n_{i\uparrow}$ describes the possibility for the site $i$ to have only one electron with spin up, $n_{i\downarrow}$ is the possibility for the site to have one electron with spin down and $n_{i2}$ is the possibility to have two electrons. The situation when site $i$ has no electrons corresponds to $1-n_{i\uparrow}-n_{i\downarrow}-n_{i2}$. It means that each site can play both roles of $A$-type site or $B$-type site. The current between sites $i$ and $j$ can be divided into four parts corresponding to these roles
\begin{equation}\label{cur-div}
J_{ji} = J_{ji}^{AA} + J_{ji}^{AB} + J_{ji}^{BA} + J_{ji}^{BB},
\end{equation}
\begin{equation}\label{JAB}
\begin{array}{l}
J_{ji}^{AA} = W_{ji}^{XX} \overline{ n_{j\uparrow} (1-n_{i\uparrow}-n_{i\downarrow}-n_{i2}) }
-W_{ij}^{XX} \overline{n_{i\uparrow} (1-n_{j\uparrow}-n_{j\downarrow}-n_{j2})}
\\
J_{ji}^{AB} = W_{ji}^{AB} \overline{n_{j\uparrow} n_{i\downarrow}}
-W_{ij}^{BA} \overline{ n_{i2} (1-n_{j\uparrow}-n_{j\downarrow}-n_{j2})}
\\
J_{ji}^{BA} = W_{ji}^{BA} \overline{ n_{j2}(1-n_{i\uparrow}-n_{i\downarrow}-n_{i2})} - W_{ij}^{AB}
\overline{n_{i\uparrow} n_{j\downarrow}}
\\
J_{ji}^{BB} = W_{ji}^{XX} \overline{ n_{j2} n_{i\downarrow}} - W_{ij}^{XX} \overline{ n_{i2}  n_{j\downarrow}}
\end{array}
\end{equation}
Here $J_{ji}$  denotes the current carried by spin-up electrons. Due to the spin inversion symmetry assumed in this study it is equal to the current carried by the spin-down electrons and to the half of the total current between sites $i$ and $j$. In the equilibrium each term in the current $J_{ij}$ is equal to zero. The averaging is performed over all right-hand side of the equations, therefore pair correlations contribute to the currents described by Eqs. (\ref{cur-div},\ref{JAB}).

The electron transition rates $W_{ij}$ acquired additional upper indexes in (\ref{JAB}) corresponding to the roles of sites $i$ and $j$.  The transition rate depends on the number of hopping electron (the first or the second electron) and on the state of final site of the hop (a hop to an empty site or to a single occupied one). For example
$W_{ij}^{AB}$ is the hopping probability for electron to hop from site $i$ to site $j$, when the site $i$ plays the role of $A$-type site and the site $j$ plays the role of $B$-type site. In other words it is the probability to hop for the first electron from site  $i$ to the single-occupied site $j$. $W_{ij}^{XX}$ can stand for both $W_{ij}^{AA}$ and $W_{ij}^{BB}$. $W_{ij}^{XX}$, $W_{ij}^{AB}$ and $W_{ij}^{BA}$ have different dependence on the Hubbard energy $E_{hub}$.
\begin{equation}\label{WXX}
\begin{array}{l}
W_{ij}^{XX} = \omega_0  \exp\left( - \frac{2r_{ij}}{a} - \frac{E_j-E_i + |E_j-E_i|}{2 k_B T} \right) ,\\
W_{ij}^{AB} =  \omega_0  \exp\left( - \frac{2r_{ij}}{a} - \frac{E_j+E_{hub}-E_i + |E_j+E_{hub}-E_i|}{2 k_B T} \right) ,\\
W_{ij}^{BA} = \omega_0  \exp\left( - \frac{2r_{ij}}{a} - \frac{E_j-E_i-E_{hub} + |E_j-E_i-E_{hub}|}{2 k_B T} \right) .\\
\end{array}
\end{equation}
Here $r_{ij}$ is the distance between sites $i$ and $j$, $a$ is the localization length, $E_i$ is the energy of the single-occupation of site $i$. $\omega_0$ is the common pre-exponent of all the hopping rates.
When the Hubbard energy is zero, $W_{ij}^{XX} = W_{ij}^{AB} = W_{ij}^{BA}$.

There are five independent pair correlations related to sites $i$ and $j$. We present them in terms of the correlation vector $\vec{\nu}_{ij}$
\begin{equation}\label{nu-cor}
\vec{\nu}_{ij} =
\left(
\begin{array}{c}
\nu_{ij,\uparrow\downarrow} - \nu_{ij,\uparrow\uparrow}\\ \nu_{ij,\uparrow\uparrow} + \nu_{ij,\uparrow\downarrow} \\ \nu_{ij,2\uparrow}\\ \nu_{ij,\uparrow2}\\ \nu_{ij,22}
\end{array}
\right) =
\left(
\begin{array}{c}
s_{ij}\\ c_{ij}\\ \nu_{ij,2\uparrow}\\ \nu_{ij,\uparrow2}\\ \nu_{ij,22}
\end{array}
\right)
\end{equation}
Here we introduced notation $\nu_{ij,xy} = \overline{n_{i,x}n_{j,y}} - \overline{n}_{i,x} \overline{n}_{j,y}$ where $x$ and $y$ can describe one of the following states of a site:  states with one electron $\uparrow$ and $\downarrow$ and the double occupied state $2$. The first two correlations $s_{ij}$ and $c_{ij}$ are discussed in the main text. The last three correlations become independent when  the sites $i$ and $j$ can play both roles of $A$-site or $B$-site. When the roles of the sites are fixed in the limit of large Hubbard energy these new correlations are either zero or they are proportional to $c_{ij}$. Naturally $\nu_{ij,2 \uparrow} = 0$ when the site $i$ is of $A$-type and $\nu_{ij,2\uparrow} = -c_{ij}$ when site $i$ is a $B$-type site. $\nu_{ij,22}$ is equal to zero when any of sites $i$ and $j$ is $A$-site. When both of them are $B$-sites it is equal to $2c_{ij}$.

When we consider correlations in a pair of sites $i-j$, all other sites should be considered as uncorrelated from $i$ and $j$. It is useful to write the hopping rate from the site $i$ to some site $k$, other than $j$, in the unified form. However it is important which of the electrons hops: the second electron from the double occupied site $i$ or the first electron. Therefore we introduce
\begin{equation}\label{Ups->}
\Upsilon_{i\rightarrow k}^A = W_{ik}^{AA}(1-2\overline{n}_{k \uparrow} - \overline{n}_{k2}) + W_{ik}^{AB} \overline{n}_{k \uparrow}, \quad
\Upsilon_{i\rightarrow k}^B = W_{ik}^{BA}(1-2\overline{n}_{k \uparrow} - \overline{n}_{k2}) + W_{ik}^{BB} \overline{n}_{k \uparrow}.
\end{equation}
In the similar way we introduce the rates for electron to come from the third site $k$ to site $i$ of the pair
\begin{equation}\label{Ups<-}
\Upsilon_{i\leftarrow k}^A = W_{ki}^{AA} \overline{n}_{k \uparrow} + W_{ki}^{BA} \overline{n}_{k2}, \quad
\Upsilon_{i\leftarrow k}^B = W_{ki}^{AB} \overline{n}_{k \uparrow} + W_{ki}^{BB} \overline{n}_{k2}.
\end{equation}
The hopping rates $\Upsilon$ are given for up projection of the spin. Similar rates for spin-down electrons can be expressed with the same equations (\ref{Ups->},\ref{Ups<-}) due to the spin inversion symmetry and the neglecting of the triple correlations.

Let us write the kinetic equation for the correlations $\vec{\nu}_{ij}$. We start from the correlation $\nu_{ij,22}$. To do this we write the rate equations for $\overline{n_{i2}n_{j2}}$ and for averaged filling numbers $\overline{n}_{i2}$ and $\overline{n}_{j2}$.
\begin{equation}\label{nu22-1}
\frac{d  \overline{n_{i2}n_{j2}}}{dt} =\sum_k \left( \Upsilon_{i \leftarrow k}^B 2 \overline{n_{i \uparrow} n_{j2}} - 2\Upsilon_{i \rightarrow k}^B \overline{n_{i2} n_{j2}} \right) + \sum_m \left( \Upsilon_{j \leftarrow m}^B 2 \overline{n_{i2} n_{j \uparrow}} - 2\Upsilon_{j \rightarrow m}^B \overline{n_{i2} n_{j2}} \right)
\end{equation}
\begin{equation}\label{nu22-2}
\frac{d \overline{n}_{i2}}{dt} = J_{ji}^{AB} + J_{ji}^{BB} + \sum_k 2\Upsilon_{i\leftarrow k}^B \overline{n}_{i \uparrow} - 2\Upsilon_{i\rightarrow k}^B \overline{n}_{i2}
\end{equation}
\begin{equation}\label{nu22-3}
\frac{d \overline{n}_{j2}}{dt} = -J_{ji}^{BA} - J_{ji}^{BB} + \sum_m 2\Upsilon_{j\leftarrow m}^B \overline{n}_{j\uparrow} - 2\Upsilon_{j\rightarrow m}^B \overline{n}_{j2}
\end{equation}
Let us note that equations (\ref{nu22-1}-\ref{nu22-3}) should in principle contain additional terms related to the correlations $i-k$ and $j-m$. We neglect this terms with the assumption of the absence of long-range correlations made in the main text. Of course, these terms can be important in other situation, for example for the calculation of the currents $J_{ki}$ and $J_{mj}$.

From the expressions (\ref{nu22-1}-\ref{nu22-3}) we derive the rate equation for $\nu_{ij,22}$
\begin{equation}\label{kin-nu-2-2}
\frac{d\nu_{ij,22}}{dt} = \overline{n}_{i2}(J_{ji}^{BA} + J_{ji}^{BB}) - \overline{n}_{j2}(J_{ji}^{AB} + J_{ji}^{BB}) -
\end{equation}
$$
- 2\left( \sum_k \Upsilon_{i\rightarrow k}^B + \sum_m \Upsilon^B_{j\rightarrow m} \right) \nu_{ij,22} +\sum_k 2\Upsilon_{i \leftarrow k}^B \nu_{ij,\uparrow 2} + \sum_m 2 \Upsilon_{j \leftarrow m}^B \nu_{ij,2\uparrow}.
$$
Let us note that when the site roles are fixed, $\nu_{ij,22}$ can be different from zero only when both sites are $B$-type. In this case $J_{ji}^{AB} = J_{ji}^{BA} = 0$ and $2\nu_{ij,\uparrow2} = -\nu_{ij,22}$. The equation (\ref{kin-nu-2-2}) reproduces the equation for $2c_{ij}$ in this case.

With the similar calculations we get
\begin{equation}\label{kin-nu-2-up}
\frac{d \nu_{ij,2\uparrow}}{dt} = J_{ji}^{BB}(1- \overline{n}_{i2} - 2\overline{n}_{j\uparrow}) - \overline{n}_{i2} J_{ji}^{BA} + J_{ji}^{AB}(\overline{n}_{i2} - 2\overline{n}_{j\uparrow}) + J_{ji}^{AA} \overline{n}_{i2} + \sum_k \left( \Upsilon_{i \leftarrow k}^B c_{ij}  - 2\Upsilon_{i \rightarrow k}^B \nu_{ij,2\uparrow} \right) +
\end{equation}
$$
 + \sum_{m} \Upsilon_{j\rightarrow m}^B \nu_{ij,22} - \left( \Upsilon_{j \rightarrow m}^{A} + \Upsilon_{j \leftarrow m}^B \right)\nu_{ij,2\uparrow}
 - \Upsilon_{j \leftarrow m}^A (2\nu_{ij,2\uparrow} + \nu_{ij,22}).
$$
The equation for $\nu_{ij, \uparrow 2}$ is similar to (\ref{kin-nu-2-up}) with substitution $i\leftrightarrow j$.

Finally we give the generalized equations for the correlations $s_{ij}$ and $c_{ij}$ mentioned in the main text
\begin{equation} \label{kin-s}
\frac{d s_{ij}}{dt} = J_{ji}^{BA} - J_{ji}^{AB} - \left[ \frac{1}{\tau_{spin}} + \sum_k \left( \Upsilon_{i\rightarrow k}^A  + \Upsilon_{i \leftarrow k}^B \right) + \sum_m \left( \Upsilon_{j \rightarrow m}^A + \Upsilon_{j \leftarrow m}^B \right) \right]s_{ij},
\end{equation}
\begin{equation}\label{kin-d}
\frac{d}{dt}c_{ij} = J_{ji}^{AA}(2\overline{n}_{i\uparrow} - 2\overline{n}_{j\uparrow}) + J_{ji}^{BA}(1-2\overline{n}_{i\uparrow} - 2\overline{n}_{j\uparrow}) + J_{ji}^{AB}(2\overline{n}_{i\uparrow} + 2\overline{n}_{j\uparrow} - 1) + J_{ji}^{BB}(2\overline{n}_{j\uparrow} - 2\overline{n}_{i\uparrow}) -
\end{equation}
$$
- \sum_k \left[ 2\Upsilon_{i \leftarrow k}^A (c_{ij} + \nu_{ij,2\uparrow} ) + (\Upsilon_{i \rightarrow k}^A + \Upsilon_{i \leftarrow k}^B) c_{ij}
- 2\Upsilon_{i\rightarrow k}^B \nu_{ij,2\uparrow} \right] -
$$
$$
- \sum_m \left[ 2\Upsilon_{j \leftarrow m}^A (c_{ij} + \nu_{ij,\uparrow 2} ) + (\Upsilon_{j \rightarrow m}^A + \Upsilon_{j \leftarrow m}^B) c_{ij}
- 2\Upsilon_{j\rightarrow m}^B \nu_{ij,\uparrow 2} \right].
$$
The equations (\ref{kin-nu-2-2}-\ref{kin-d}) express the same physics as the equations for correlations $s_{ij}$ and $c_{ij}$ in the limit of large Hubbard energy. All the correlations  are generated by the applied currents and relax due to hopping to the third sites. The relaxation of different correlations is however not independent. Correlations can produce each other in the process of relaxation.

The effect of the correlations on the currents is described by Eq. (\ref{JAB}). This equation includes joint averaging of the products of different filling numbers. This products can be separated into parts corresponding to averaged filling numbers $\overline{n}_{ix}$, $\overline{n}_{jy}$ and correlations $\vec{\nu}_{ij}$. Therefore the situation is similar to the limit of the large Hubbard energy.
\vspace{1cm}

\hspace{5cm} \textbf{Matrix form of the kinetic equations.}
\vspace{1cm}

It is useful to write the generalized kinetic equation for the pair correlations and their effect on the currents in the matrix form.
\begin{equation}\label{cor-mat1}
\frac{d}{dt} \nu_{ij}^{\alpha} = G_{ij}^{\alpha X} J_{ji}^{X} - R_{ij}^{\alpha\beta}\nu_{ij}^{\beta}.
\end{equation}
Here $\alpha$ and $\beta$ enumerate components of the vector $\vec{\nu}$. Index $X$ corresponds to one of the four currents $J_{ji}^X$ and can have one of the four values $AA$, $AB$, $BA$ or $BB$. The generation matrix $G_{ij}^{\alpha X}$ and the relaxation matrix $R_{ij}^{\alpha\beta}$ are defined from Eqs. (\ref{kin-nu-2-2} --- \ref{kin-d})
\begin{equation} \label{Gmat}
\widehat{G}_{ij} = \left(
\begin{array}{l||c|c|c|c}
\,  & AA & AB & BA & BB \\
\hline \hline
s_{ij}  & 0 & -1 & 1 & 0  \\
\hline
c_{ij} & 2(\overline{n}_{i\uparrow} - \overline{n}_{j\uparrow}) & 2\overline{n}_{i\uparrow} + 2\overline{n}_{j\uparrow} -1 & 1-2\overline{n}_{i\uparrow} - 2\overline{n}_{j\uparrow} & 2(\overline{n}_{j\uparrow} - \overline{n}_{i\uparrow}) \\
\hline
\nu_{ij, 2\uparrow} & \overline{n}_{i2} & \overline{n}_{i2} - 2 \overline{n}_{j\uparrow} & -\overline{n}_{i2} & 1- \overline{n}_{i2}- 2\overline{n}_{j\uparrow}  \\
\hline
\nu_{ij, \uparrow 2} & -\overline{n}_{j2} & \overline{n}_{j2} & 2\overline{n}_{i\uparrow} - \overline{n}_{j2} & -1+\overline{n}_{j2} + 2\overline{n}_i \\
\hline
\nu_{ij,22} & 0 & -\overline{n}_{j2} & \overline{n}_{i2} & \overline{n}_{i2} - \overline{n}_{j2}
\end{array}
\right)
\end{equation}
The relaxation matrix $R_{ij}^{\alpha\beta}$ can be given as a sum of contributions related to all the sites other than $i$ and $j$
\begin{equation} \label{Rmatr}
R_{ij,\alpha \beta} = \frac{1}{\tau_s} \delta_{\alpha,s}\delta_{\beta, s} + \sum_k R_{ij,\alpha\beta}^{(k)} + \sum_m R_{ij,\alpha\beta}^{(m)},
\end{equation}
\begin{equation}
\widehat{R}_{ij}^{(k)} = \left(
\begin{array}{c|c|c|c|c}
 \Upsilon_{i \rightarrow k}^{A} + \Upsilon_{i \leftarrow k}^{B} & 0 & 0 & 0 & 0  \\
\hline
 0 & 2\Upsilon_{i \leftarrow k}^A + \Upsilon_{i \rightarrow k}^A + \Upsilon_{i \leftarrow k}^B  & 2\Upsilon_{i \leftarrow k}^A - 2 \Upsilon_{i \rightarrow k}^B & 0 & 0 \\
\hline
 0 & -\Upsilon_{i \leftarrow k}^B & 2 \Upsilon_{i \rightarrow k}^B & 0 & 0  \\
\hline
 0 & 0 & 0 & \Upsilon_{i \rightarrow k}^A + \Upsilon_{i \leftarrow k}^B + 2\Upsilon_{i \leftarrow k}^A & \Upsilon_{i \leftarrow k}^A - \Upsilon_{i \rightarrow k}^B \\
\hline
 0 & 0 & 0 & -2 \Upsilon_{i \leftarrow k}^B & 2 \Upsilon_{i \rightarrow k}^B
\end{array}
\right),
\end{equation}
\begin{equation}
\widehat{R}_{ij}^{(m)} =
\end{equation}
$$
=\left(
\begin{array}{c|c|c|c|c}
 \Upsilon_{j \rightarrow m}^{A} + \Upsilon_{j \leftarrow m}^{B} & 0 & 0 & 0 & 0  \\
\hline
 0 & 2\Upsilon_{j \leftarrow m}^A + \Upsilon_{j \rightarrow m}^A + \Upsilon_{j \leftarrow m}^B  & 0 & 2\Upsilon_{j \leftarrow m}^A - 2\Upsilon_{j \rightarrow m}^B & 0 \\
\hline
 0 & 0 & \Upsilon_{j \rightarrow m}^A + \Upsilon_{j \leftarrow m}^B + 2\Upsilon_{j \leftarrow m}^A & 0 &
\Upsilon_{j \leftarrow m}^A - \Upsilon_{j \rightarrow m}^B  \\
\hline
 0 &- \Upsilon_{j \leftarrow m}^B & 0 & 2\Upsilon_{j \rightarrow m}^B & 0 \\
\hline
 0 & 0 & -2\Upsilon_{j \leftarrow m}^B & 0 & 2 \Upsilon_{j \rightarrow m}^B
\end{array}
\right).
$$

The current generation due to the correlations can also be given in the matrix form
\begin{equation} \label{J-mat}
J_{ji}^X = J_{ji,0}^X + {\cal W}_{ij}^{X\alpha} \nu_{ij}^{\alpha}.
\end{equation}
Here the current generation matrix ${\cal W}_{ij}^{X\alpha}$ is
\begin{equation} \label{Wmat}
{\cal W}_{ij}^{X \alpha} =
\left(
\begin{array}{l||c|c|c|c|c}
\, & s_{ij} & c_{ij} & \nu_{ij,2\uparrow} & \nu_{ij, \uparrow,2} & \nu_{ij,22} \\
\hline \hline
J_{ji}^{AA} & 0 & -W_{ji}^{AA} + W_{ij}^{AA} &  - W_{ji}^{AA} & W_{ij}^{AA} & 0 \\
\hline
J_{ji}^{AB} & W_{ji}^{AB} /2 & W_{ji}^{AB}/2 & 2W_{ij}^{BA} & 0 & W_{ij}^{BA} \\
\hline
J_{ji}^{BA} & - W_{ij}^{AB}/2 & - W_{ij}^{AB} /2 & 0 & -2W_{ji}^{BA} & - W_{ji}^{BA} \\
\hline
J_{ji}^{BB} & 0 & 0 & - W_{ij}^{BB} & W_{ji}^{BB} & 0
\end{array}
\right).
\end{equation}
 $J_{ji,0}^X$ are the currents that are calculated without correlations, i.e. with equation (\ref{JAB}) where all the averaged products of the filling numbers are decoupled into the products of averaged filling numbers
\begin{equation}\label{JAB0}
\begin{array}{l}
J_{ji,0}^{AA} = W_{ji}^{XX} \overline{n}_{j\uparrow} (1-\overline{n}_{i\uparrow}-\overline{n}_{i\downarrow}-\overline{n}_{i2})
-W_{ij}^{XX} \overline{n}_{i\uparrow} (1-\overline{n}_{j\uparrow}-\overline{n}_{j\downarrow}-\overline{n}_{j2}),
\\
J_{ji,0}^{AB} = W_{ji}^{AB} \overline{n}_{j\uparrow} \overline{n}_{i\downarrow}
-W_{ij}^{BA} \overline{ n}_{i2} (1-\overline{n}_{j\uparrow}-\overline{n}_{j\downarrow}-\overline{n}_{j2}),
\\
J_{ji,0}^{BA} = W_{ji}^{BA} \overline{ n}_{j2}(1-\overline{n}_{i\uparrow}-\overline{n}_{i\downarrow}-\overline{n}_{i2}) - W_{ij}^{AB}
\overline{n}_{i\uparrow} \overline{n}_{j\downarrow},
\\
J_{ji,0}^{BB} = W_{ji}^{XX} \overline{ n}_{j2} \overline{n}_{i\downarrow} - W_{ij}^{XX} \overline{ n}_{i2}  \overline{n}_{j\downarrow}.
\end{array}
\end{equation}

Finally the system of equations should include the kinetic equations for $\overline{n}_{i\uparrow}$ and $\overline{n}_{i2}$.
\begin{equation} \label{kin-n2}
\frac{d \overline{n}_{i2}}{dt} = 2\sum_j J_{ji}^{AB} + J_{ji}^{BB},
\end{equation}
\begin{equation} \label{kin-n-up}
\frac{d \overline{n}_{i\uparrow}}{dt} = \sum_j J_{ji}^{AA} + J_{ji}^{BA} -  J_{ji}^{AB} - J_{ji}^{BB}.
\end{equation}

Together equations (\ref{cor-mat1} --- \ref{kin-n-up}) form a closed system of equations that should be solved to describe the hopping transport when only pair correlations are considered.

\vspace{1cm}

\hspace{5cm} \textbf{Linearization of the kinetic equations.}
\vspace{1cm}

In the main text we show that in the limit of large Hubbard energy the linearization of the kinetic equations yields the effective resistor network that generalize the Miller-Abrahams network. Here we make the linearization in the general case of arbitrary $E_{hub}$.

The number of the effective charge conservation laws (\ref{kin-n2},\ref{kin-n-up}) associated with each site $i$ is two. It is related to the fact that there is no re-distribution between the first and the second electrons on the site --- if some site $i$ acquired the second electron, it should first loose this second electron due to some hop and only than it can loose its first electron. It means that each site should be represented by two nodes on the effective scheme and ascribed by two electro-chemical potentials $\varphi_i^A$ and $\varphi_i^B$ corresponding to the first and the second electron. The electrical potentials for both electrons are of course equal (they are determined by the electric field and site coordinates) but the chemical potentials can be different out of the equilibrium.

The ``non-correlated'' currents $J_{ji,0}^X$ are governed by these potentials and the analog of Miller-Abrahams resistors
\begin{equation}\label{linJ0}
J_{ji,0}^{PQ} = \frac{\varphi_j^P - \varphi_i^Q}{R_{ji,0}^{PQ}}.
\end{equation}
Here indexes $P$ and $Q$ can have one of the two values $A$ and $B$ (the pair of indexes $PQ$ correspond to one index $X$). $R_{ji,0}^{PQ}$ is the analog of Miller-Abrahams resistor corresponding to the part of the current $J_{ji,0}^{PQ}$. It can be estimated as
\begin{equation}
R_{ji,0}^{PQ} = R_0 \exp \left(\frac{2 r_{ij}}{a} + \frac{|E_j^P| + |E_i^Q| + |E_j^P - E_i^Q|}{2 k_B T}\right).
\end{equation}
It is similar to the ordinary expression of the Miller-Abrahams resistor where the energy $E_i^Q$ can stand for the energy of single or double occupation $E_i^A = E_i$, $E_i^B = E_i + E_{hub}$.

The actual currents are related to $J_{ji,0}^{PQ}$ as described before
\begin{equation}\label{linJ}
J_{ji}^{PQ} = \left[ \widehat{1} - \widehat{\cal W}_{ij} \widehat{R}_{ij}^{-1} \widehat{G}_{ij} \right]^{-1}_{PQ,P'Q'} J_{ji,0}^{P'Q'}.
\end{equation}
Here $\widehat{\cal W}_{ij}$, $\widehat{R}_{ij}$ and $\widehat{G}_{ij}$ are the matrixes described by the equations (\ref{Wmat}), (\ref{Rmatr}) and (\ref{Gmat}) correspondingly. The substitution of the equations (\ref{linJ0}) and (\ref{linJ}) to (\ref{kin-n2}) and (\ref{kin-n-up}) leads to the generalized system of the Kirchhoff equations
\begin{equation}\label{Kirh1}
\sum_j \left[ \widehat{1} - \widehat{\cal W}_{ij} \widehat{R}_{ij}^{-1} \widehat{G}_{ij} \right]^{-1}_{AA,PQ} \frac{\varphi_j^{P} -\varphi_i^{Q} }{R_{ji}^{PQ}} + \left[ \widehat{1} - \widehat{\cal W}_{ij} \widehat{R}_{ij}^{-1} \widehat{G}_{ij} \right]^{-1}_{BA,PQ} \frac{\varphi_j^{P} -\varphi_i^{Q} }{R_{ji}^{PQ}} = 0
\end{equation}
\begin{equation}\label{Kirh2}
\sum_j \left[ \widehat{1} - \widehat{\cal W}_{ij} \widehat{R}_{ij}^{-1} \widehat{G}_{ij} \right]^{-1}_{AB,PQ} \frac{\varphi_j^{P} -\varphi_i^{Q} }{R_{ji}^{PQ}} + \left[ \widehat{1} - \widehat{\cal W}_{ij} \widehat{R}_{ij}^{-1} \widehat{G}_{ij} \right]^{-1}_{BB,PQ} \frac{\varphi_j^{P} -\varphi_i^{Q} }{R_{ji}^{PQ}} = 0
\end{equation}
The equations (\ref{Kirh1}) and (\ref{Kirh2}) form the linear system that can be solved to find all the potentials $\varphi_i^Q$ as in the case of usual system of the Kirchhoff equations. However (\ref{Kirh1}) and (\ref{Kirh2}) cannot be interpreted as an equivalent scheme of resistors. Naturally, the difference of electro-chemical potentials $\varphi_j^B-\varphi_i^A$ not only lead to the current $J_{ji}^{BA}$ but due to the correlations it also leads to the currents $J_{ji}^{AA}$, $J_{ji}^{AB}$ and $J_{ji}^{BB}$. The behavior of system in the linear regime is therefore more complex than the one of a scheme of ordinary resistances.

\vspace{1cm}

\hspace{5cm} \textbf{Numerical solution of the kinetic equations.}
\vspace{1cm}

We have solved kinetic equations (\ref{cor-mat1} --- \ref{kin-n-up}) using the standard Euler method. We apply the standard periodic boundary conditions in the presence of the constant electric field. In order to eliminate the influence of the finite size effects we performed the calculation of the magnetoresistance for different system size. In Fig. 1 we plot averaged over 100 configurations of disorder magnetoresistance calculated for
$32\times32$ and $64\times64$ systems. As it is seen from the figure there is marginal difference between these calculations. Therefore we can conclude that system sizes $32\times32$ and $64\times64$ are big enough in order to neglect the finite size effects.
This is demonstrated in Fig.2, where the finite size scaling for the magnetoresistance  is plotted as a function of the inverse system size. Clearly the finite size effects are smaller then 1\%.
\begin{figure}[htbp]
    \centering
        \includegraphics[width=0.5\textwidth]{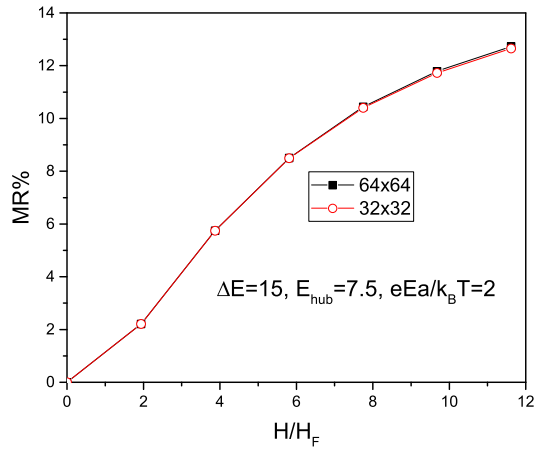}
        \caption{The dependence of the magnetoresistance on magnetic field, calculated
         for different system size $32\times32$ and $64\times64$. }
    \label{fig:size}
\end{figure}
\begin{figure}[htbp]
    \centering
        \includegraphics[width=0.5\textwidth]{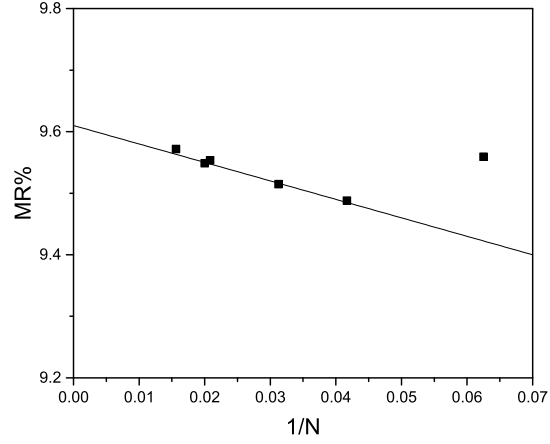}
        \caption{The dependence of the magnetoresistance in the magnetic field 
        $H/H_F=11.6$ on the inverse system size $1/N$. }
    \label{fig:scaling}
\end{figure}

Similarly we averaged the calculated magnetoresistance over 100 different random energy realization. In order to demonstrate the effect of averaging we have calculated the average magnetoresistance for different number of random energy realizations. In Fig.3 we plot the average magnetoresistance averaged over 100 and 200 random energy realization. As it is seen from the figure the results of different averaging are almost identical. Therefore we can conclude that averaging over 100 different realization of disorder is good enough to produce reliable results for magnetoresistance.

\begin{figure}[htbp]
    \centering
        \includegraphics[width=0.5\textwidth]{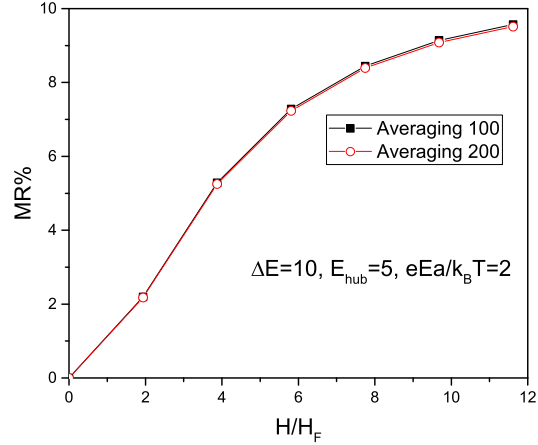}
        \caption{The dependence of the magnetoresistance on magnetic field, calculated
         for different number of random energy realizations. }
    \label{fig:average}
\end{figure}

\end{document}